\documentclass[conference]{IEEEtran}

\ifCLASSINFOpdf
            \else
                  \fi

\usepackage{graphicx}
\graphicspath{{figures/}}
\DeclareGraphicsExtensions{.jpg,.png}
\usepackage{etoolbox}
\usepackage[caption=false,font=footnotesize]{subfig}

\usepackage{multirow,makecell}

\usepackage{framed}
\usepackage{float}

\usepackage[nodayofweek]{datetime}
\usepackage{listings}

\usepackage{array}
\usepackage{rotating}
\usepackage{adjustbox}
\usepackage{tabularx}
\usepackage[framemethod=default]{mdframed}
\usepackage{tikz}
\usepackage{pgfplots, pgfplotstable}
\usetikzlibrary{patterns}
\usepackage{booktabs}

\usepackage{rotating}
\usepackage{framed}
\usepackage{float}
\newfloat{Protocol}{t}{struct}

\usepackage{makecell}
\usepackage{multirow}
\usepackage{comment}

\usepackage[ruled,vlined]{algorithm2e}
\usepackage{algpseudocode}
\usepackage{tikz}
\usepackage{pgfplots}
\usetikzlibrary{patterns}

\usepackage{amssymb}
\usepackage{listings}
\usepackage{pifont}
\makeatletter
\def\blfootnote{\xdef\@thefnmark{}\@footnotetext}
\makeatother

\usepackage{pgfplotstable}
\usepackage{filecontents}

\usepackage{array}
\usepackage{xcolor,colortbl}
\usepackage[hidelinks,pdftitle={Title},pdfauthor={anonymous},unicode]{hyperref}
\usepackage{adjustbox}
\usepackage{fancyvrb}
\usepackage{balance}
\usepackage{mathtools}
  
\usepackage{blindtext}

\usepackage{xspace}

\newcommand{\ie}{\textit{i.e.,}\xspace}
\newcommand{\eg}{\textit{e.g.,}\xspace}

\newcounter{protocol}[section]

\newcolumntype{B}[2]{    >{\adjustbox{angle=#1,lap=\width-(#2)}\bgroup}    l    <{\egroup}}

\newcolumntype{Z}[2]{    >{\adjustbox{angle=#1,lap=\width-(#2)}\bgroup}    l    <{\egroup}}

\newcolumntype{L}[1]{>{\raggedright\let\newline\\\arraybackslash\hspace{0pt}}m{#1}}
\newcolumntype{C}[1]{>{\centering\let\newline\\\arraybackslash\hspace{0pt}}m{#1}}
\newcolumntype{R}[1]{>{\raggedleft\let\newline\\\arraybackslash\hspace{0pt}}m{#1}}

\definecolor{semi-light-gray}{gray}{0.7}
\definecolor{light-gray}{gray}{0.8}

\begin{document}
\title{Secure Client and Server Geolocation Over the Internet}

\author{\IEEEauthorblockN{AbdelRahman Abdou}
\IEEEauthorblockA{Carleton University, Ottawa, Canada\\
ETH Z\"{u}rich, Switzerland\\
abdoua@inf.ethz.ch}
\and
\IEEEauthorblockN{Paul C.\ van Oorschot}
\IEEEauthorblockA{School of Computer Science\\
Carleton University, Ottawa, Canada\\
paulv@scs.carleton.ca}
}

\maketitle

\begin{abstract}
In this article, we provide a summary of recent efforts towards achieving Internet geolocation securely, \ie without allowing the entity being geolocated to cheat about its own geographic location. Cheating motivations arise from many factors, including impersonation (in the case locations are used to reinforce authentication), and gaining location-dependent benefits. In particular, we provide a technical overview of Client Presence Verification (CPV) and Server Location Verification (SLV)---two recently proposed techniques designed to verify the geographic locations of clients and servers in realtime over the Internet. Each technique addresses a wide range of adversarial tactics to manipulate geolocation, including the use of IP-hiding technologies like VPNs and anonymizers, as we now explain.
\end{abstract}

\IEEEpeerreviewmaketitle

\blfootnote{A version of this paper appeared in the USENIX ;login: magazine, Spring 2018 issue. This is the authors' copy for personal use.}

\section{Introduction}

Internet Geolocation is the process of determining the geographic location of an Internet-connected device. Secure geolocating of a web client (a client visiting a website) is useful for location-aware authentication, location-aware access control, location-based online voting, location-based social networking, and fraud reduction.

From the client's perspective, geolocating the remote server can provide higher assurance to the server's identity, and justify conducting certain sensitive transactions, \eg those requiring certain privacy measures or requiring data sovereignty~\cite{peterson2011position}. Independent of server and client geolocation, geolocating network intermediate systems (\eg routers) can also be useful for purposes such as monitoring~\cite{huffaker2014drop} and network mapping~\cite{csoma2017measuring}.

This article provides a technical overview on two recent approaches for secure location verification of clients and servers over the Internet, respectively CPV and SLV. Both techniques are based on network measurements, where delays are measured from trusted network nodes dubbed \emph{verifiers}, and are analyzed in realtime to verify physical presence inside a prescribed geographic region. We explain the threat model of both techniques, how they mitigate against known adversarial tactics, how they adapt to various network dynamics, and what distinguishes them from other geolocation approaches.

\definecolor{light-gray}{gray}{0.95}
\lstdefinelanguage{JavaScript}{
  keywords={typeof, new, true, false, catch, function, return, null, catch, switch, var, if, in, while, do, else, case, break},
  keywordstyle=\color{blue}\bfseries,
  ndkeywords={class, export, boolean, throw, implements, import, this},
  ndkeywordstyle=\color{darkgray}\bfseries,
  identifierstyle=\color{black},
  sensitive=false,
  comment=[l]{//},
  morecomment=[s]{/*}{*/},
  commentstyle=\color{purple}\ttfamily,
  stringstyle=\color{red}\ttfamily,
  morestring=[b]',
  morestring=[b]"
}

\lstset{   backgroundcolor=\color{light-gray},  basicstyle=\scriptsize\ttfamily,  breakatwhitespace=false,         	  breaklines=true,                 	  captionpos=b,                    	  commentstyle=\color{ForestGreen},    	  extendedchars=true,             	  frame=single,                    	  keepspaces=true,                 	  keywordstyle=\color{blue},       	  language=JavaScript,					  numbersep=5pt,                   	    numberstyle=\tiny\color{gray}, 		  stepnumber=1,                    	  rulecolor=\color{black},         	  showspaces=false,                	  showstringspaces=false,          	  showtabs=false,                  	  tabsize=2,                       	    escapeinside={(*@}{@*)},
  stringstyle=\color{orange},
  showstringspaces=false,
}

\section{Geolocation Background}

This section provides a background on how Internet geolocation works, and the security limitations associated with common practices. Many academic geolocation methods have been proposed, but there have been very limited deployment in practice. As of this writing, most of the geolocation conducted in practice relies on the clients' IP address or GPS coordinates of hand-held devices, as explained below. 

\subsection{Geolocation in Practice}

\begin{figure}
\centering
\includegraphics[scale=0.3]{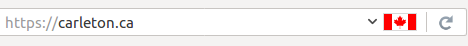}
\includegraphics[scale=0.3]{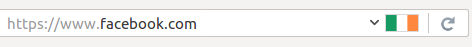}
\includegraphics[scale=0.3]{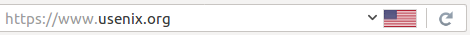}
\caption{Snapshots of the \emph{Flagfox} browser extension.}
\label{fig:flagfox}
\end{figure}

There are several methods for device geolocation over the Internet. If the device belongs to a user that is acting as a web client (\ie visiting a website), the Geolocation API is a W3C standard that enables browsers to obtain location information of the device they are running on, and communicate it to a webserver. Servers request location coordinates using Javascript as follows: 
\begin{lstlisting}
if(navigator.geolocation) {
		navigator.geolocation.getCurrentPosition(success, error, geoOptions);
	} else {
		console.log("Geolocation is not supported on this browser.");
	}
\end{lstlisting}
The geolocation methods a browser uses are left to the browser vendor's discretion. Most major browsers rely on the following in varying orders (\ie when one fails, the next is tried): GPS, WiFi Positioning System (WPS), IP address-based location look-ups, or cell-tower triangulation of mobile devices. The location of an IP address can be obtained from, \eg publicly available routing information or public registries, such as \texttt{whois}. Many IP location service providers (commercial and free) maintain look-up tables to instantly map IP addresses to locations. Such static \textit{tablulation} methods may take long times to reflect changes or IP address re-assignments, which occur quite often for client geolocation to be up to date (studies were conducted to confirm this~\cite{poese2011ip}). IP address based geolocation can however be reliable for benign server geolocation. \emph{Flagfox} is an example Firefox extension that visually indicates a flag of the country corresponding to the IP address resolution of the URL (see Fig~\ref{fig:flagfox}).

\begin{figure}
\centering
\includegraphics[scale=0.3]{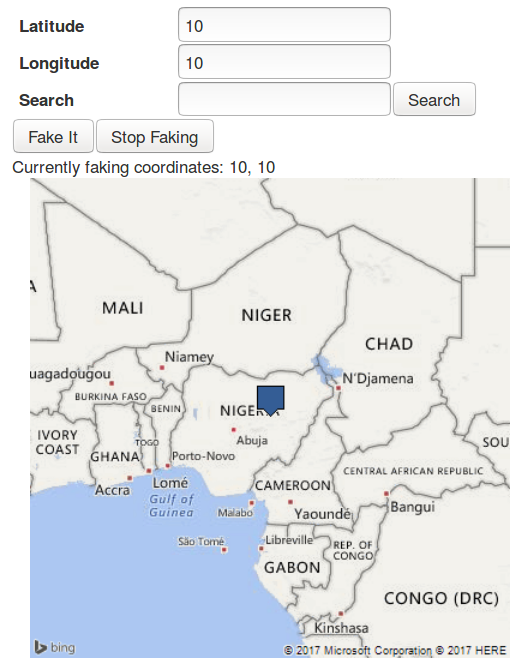}
\caption{Snapshots of the \emph{Fake Location} extension---an example browser extension allowing users to fake their locations.}
\label{fig:fakelocation}
\end{figure}

From a security perspective, none of the above techniques is resilient to adversarial manipulation. When the geolocation API is in use, the server normally makes no effort in geolocating the client device; it rather trusts the browser-communicated coordinates, which can easily be forged on the fly before being sent to the server. Example Firefox extensions that enable forgery include \emph{Fake Location} (see Fig~\ref{fig:fakelocation}) and \emph{Location Guard}; both enable a user to specify where in the world they would like to appear to be at. If the server relies on tabulation methods to geolocate the client (instead of asking the browser for its coordinates), the common practice of clients hiding their own IP addresses behind proxies and anonymizers comes into play. 

\subsection{Geolocation in the Literature}

A wide set of techniques can be used, mostly for a server to geolocate clients~\cite{muir2009internet}. These enable a server to infer a client's geographic location from, \eg hints obtained from browser-generated HTTP headers (such as preferred language or time zone). Locations can also be obtained through crowd-sourcing, \eg by interpolating a device's location from its proximity to nearby devices (phones) with known GPS locations.

Another class of Internet geolocation approaches is based on network measurements. Similar to GPS triangulations that are based on the delays between the receiver and satellites, measurement-based techniques also aim to locate devices (clients or server) by estimating their distance from landmarks in the network with known locations. These landmarks measure network delays from themselves to the device, typically identified by its IP address, and map these delays to geographic distances. The accuracy of such mapping however is not anywhere near that of mapping satellite delays to distances, and is thus the primary source of inaccuracies in such techniques. Still, measurement-based geolocation is generally considered more accurate than methods like tabulation-based geolocation.

From the security point of view, although most of the above methods are positioned as resilient to evasion, examination has shown otherwise. Delay-increasing attacks can allow an adversary to distort its perceived location~\cite{gill2010dude}. Delay-decreasing was also studied, \eg by manipulating ICMP \texttt{ping} and \texttt{traceroute} as they fail to preserve the integrity of timing measurements. Combining both attacks, an adversary can forge the calculated location to an accuracy of a few tens of kilometres relative to a target desired location~\cite{abdou2017accurate}.

\section{Client Presence Verification (CPV)}

CPV~\cite{abdou2017cpv} is a measurement-based technique designed to verify the geographic locations of web users (\emph{clients} henceforth) over the Internet. The client is assumed to be motivated to misrepresent its location, \eg to gain location-dependent benefits. CPV's design takes into consideration various adversarial location-forging tactics, including delay manipulations and IP-hiding technologies like VPNs and anonymizers. CPV does not rely fundamentally on the clients' IP addresses, nor does it determine geographic locations. Rather, it \emph{verifies} an asserted (unverified) location, typically made by a client. The client's location could be asserted using, \eg the client's GPS coordinates, the client's IP address, or even explicitly such asking the user to fill-in their street address in an online form during login.

To verify location assertions, CPV relies on an infrastructure of geographically scattered nodes, dubbed \emph{verifiers} (\eg virtual private, or cloud-based servers). The technique works as follows. When a client visits a website and asserts the geographic location from which he/she is currently browsing, three verifiers surrounding the asserted location are selected. The verifiers measure (in realtime) network delays between themselves and the client's browser, and analyze these delays to corroborate that the client is present somewhere inside the triangle determined by their (the verifiers) geographic locations. Because the verifiers cannot pinpoint where exactly the client is within the triangle, the size of the triangle is the verification granularity.

\subsection{Secure One-way Delay Estimation}

The verifiers do not measure round-trip times~(RTT) between themselves and the client. They rather estimate the smaller of the forward and reverse one-way delays~(OWDs) between each of them and the client. The larger OWD is discarded because propagation delays between two network nodes are bounded by the physical distance between them, so a smaller OWD measurement is a better representation to the geographic distance between both nodes than the larger---the larger must have been affected by other factors such as network congestion or circuitous routing.

To measure the OWD between a verifier and the client, CPV does not rely on standard OWD-estimation protocols like OWAMP (RFC 4656), as those require honest client cooperation, \eg client clock synchronization and honest reporting of delays. As such, CPV relies on the \emph{Minimum-Pairs~(MP)} protocol~\cite{abdou2015accurate}. MP requires the three verifiers, A, B, and C, to first synchronize their clocks, and pre-share cryptographic keys to ensure operational integrity.

Through Javascript, the client's browser is first directed to establish a WebSocket (RFC 6455) connection to the three Verifiers, which are chosen based on the client's asserted location. Verifier A begins by a sending a cryptographically-protected timestamp (in millisecond precision) to the client, which the browser forwards to the other two verifiers. On receiving this, verifier B calculates the propagation time from A $\xrightarrow{}$ client $\xrightarrow{}$ B, and likewise when the timestamp is received by C. Verifiers B and C then follow suit, taking turns in sending timestamps. When all three verifiers are done exchanging timestamp messages, they will have six delay values as follows:
\begin{itemize}
\itemsep0em
\item A $\xrightarrow{}$ client $\xrightarrow{}$ B.
\item A $\xrightarrow{}$ client $\xrightarrow{}$ C.
\item B $\xrightarrow{}$ client $\xrightarrow{}$ A.
\item B $\xrightarrow{}$ client $\xrightarrow{}$ C.
\item C $\xrightarrow{}$ client $\xrightarrow{}$ A.
\item C $\xrightarrow{}$ client $\xrightarrow{}$ B.
\end{itemize}
Between each pair of verifiers, \eg between \{A $\xrightarrow{}$ client $\xrightarrow{}$ B\} and \{B $\xrightarrow{}$ client $\xrightarrow{}$ A\}, the verifiers exclude the larger OWD, and solve a system of three equations simultaneously for an estimate to the smaller OWD between the client and each verifier. That is, if the smaller of the forward and reverse OWD between the client and A, B, and C respectively is $a$, $b$, $c$, then (note: $=$ sign here is used to indicate mathematical equality rather than an assignment operator):
\begin{itemize}
\itemsep0em
\item $a+b = min(AtB, BtA)$
\item $a+c = min(AtC, CtA)$
\item $b+c = min(BtC, CtB)$
\end{itemize}
where $AtB$ is the delay A $\xrightarrow{}$ target client $\xrightarrow{}$ B, and so on. Analysis of MP's accuracy showed that the protocol is likely to provide more accurate estimates to the smaller OWD than simply using half the RTT~\cite{abdou2015accurate}.

\subsection{Corroborating Presence Inside the Triangle}
\label{sec:conditions}

In order to avoid potential inaccuracies from delay-to-distance mapping, the calculated OWDs are not mapped to distances. They are rather compared to the smaller OWDs between the verifiers themselves, which are measured and updated periodically in a background process, independent of whether or not a client is currently being verified. Assuming $x = min(AB, BA)$ is the smaller of the forward and reverse OWDs between verifiers A and B \emph{directly} (not to be confused with $min(AtB, BtA)$ from the previous section), and likewise $y = min(BC, CB)$ and $z = min(AC, CA)$, then the client's asserted location is accepted as inside the triangle if:
\begin{center}
area($\triangle xab$)+area($\triangle ybc$)+area($\triangle zca$) $\leq$ area($\triangle xyz$) $+\epsilon$
\end{center}
Such that area($\triangle xab$) is the area of that triangle calculated from its side lengths $x$, $a$, and $b$. The value of $\epsilon$ is used to account for the two extra access network traversals occurring at the client when the timestamps propagate from a verifier to the client to another verifier (see Section~\ref{sec:calibration} for setting this value).

{\bf Iterative Delay Measurement.} To account for abrupt delay spikes or network irregularities, the above process of OWD calculations and comparison with those between the verifiers is iteratively repeated, \eg 20 times. If the condition is met for the majority of the conducted iterations, the location assertion is accepted.

\subsection{CPV Calibration}
\label{sec:calibration}

There are several parameters that tune CPV's reaction to events. The most important three are $\epsilon$ (in Section~\ref{sec:conditions}), $n$, which is the number of delay measurement iterations, and $\tau$, which is the fraction of those iterations that must \emph{pass} (\ie the condition is met) for the client's asserted location to be accepted. This calibration should take place before the location verification process begins. To do that, the three verifiers may use network nodes that they know as a ground truth are inside the triangle. From the network delays of these nodes, the verifiers compute values for the above mentioned three parameters, and then run CPV to verify the client's location.

\subsection{Hindering Illicit Traffic Relaying}

In an attempt to defeat geolocation, a middlebox (like a proxy server or a VPN gateway) that is physically inside the triangle can be specifically customized to filter out the verifiers' timestamps from the client's traffic and forward them to the verifiers on behalf of the client. This threat against CPV is exacerbated by the presence of numerous cheap public VPN providers whose primary service is to enable subscribers to evade geolocation technologies.

Techniques like CPV can mitigate against this by adapting known Proof-of-Work techniques~\cite{abdou2015taxing}. The verifiers generate a cryptographic client puzzle with each timestamp message, which the client's browser must solve before forwarding the message (puzzle solution and timestamp) to the other two verifiers. The puzzles must be easy to solve so that they do not (1) overwhelm the client with high processing costs, and (2) overshadow the network propagation delays.

In the case of a middlebox connected to many simultaneous (\ie cheating) clients, the middlebox will choose to either solve these puzzles on behalf of the clients, or forward them to the clients. In the latter case, the network delay between the middlebox and the client will get added to the time the verifiers observe for location verification, which results in CPV correctly detecting the client's absence from the respective triangle. It is thus in the middlebox's interest to choose the former case---solving the puzzles on behalf of the clients. However, this means that as more clients are connected, the middlebox will have to solve more puzzles. When these puzzles begin to accumulate, they will increase queueing delays, which contribute to the delays observed by the verifiers, eventually causing CPV to reject the location assertions of all middlebox-connected clients.

In this model, there are two main parameters contributing to the puzzle queueing rate at the middlebox: the puzzle difficulty and the middlebox's computational resources (\eg processor clock frequency and number of processing cores). Queueing analysis~\cite{abdou2015taxing} shows that the puzzle difficulty has a higher impact on the rate of puzzle queueing than the middlebox's computational power. This analysis suggests that this puzzle mechanism will be effective to hinder illicit middlebox relaying. 

\subsection{Evaluation Results}

CPV was evaluated using PlanetLab---a distributed testbed for Internet measurement research and network experiments---using 80 planetlab nodes in North America. Three of the nodes were selected to act as verifiers, and the remaining 77 acted as clients. Some of the 77 nodes were inside the triangle (recall: the verification triangle determined by the three verifiers), while others were outside. All 77 nodes carried out the protocol with the verifiers simultaneously to get their locations verified. Knowing the ground-truth of which nodes are inside and which are outside (the geographic locations of PlanetLab nodes are publicly disclosed on PlanetLab's website), we can count the number False Rejects (nodes inside the triangle identified by CPV as outside) and False Accepts. The process is repeated after choosing a different triangle (a different set of three nodes to act as verifiers), again counting false rejects and false accepts. In total, 34 triangles where chosen. Triangles were chosen to be nearly equilateral (physically), with inside angles ranging from 50-70 degrees (0.87-1.22 radians). The smallest triangle had an area equivalent to a circle of radius 100km, and largest 400km.

When the inside nodes are not too close to the triangle's sides (\ie away from the closest side by at least 10\% of its length), CPV resulted in a total of 1.0\% false accepts and 2.0\% false rejects~\cite{abdou2017cpv}. These results were obtained when 600 CPV iterations (see Section~\ref{sec:conditions}) were performed with each client. The results were not much different when only 100 iterations were performed, where the false accept rate increased only to 1.1\% and the false reject rate remained unchanged. However, when only 10 iterations were performed, false accepts and false rejects were at 2.1\% and 4.1\% respectively.

Testing was later repeated to assess the effect of WiFi access networks on CPV's efficacy~\cite{abdou2016wireless}. WiFi access networks often have higher delays and delay jitters. A different evaluation technique was used, as the PlanetLab infrastructure used above involved nodes connected using wired access networks. To model WiFi clients, 802.11 delay models from the literature were used to generate the last-mile delays, which were added (\ie arithmetic summation) to the delay traces collected from PlanetLab. Since higher network delays for nodes inside the triangles may result in higher false rejects, the generated 802.11 delays were only added to the the delays of inside nodes to create the most stressful testing situation. 802.11 networks employ slotted retransmissions. The delays were generated such that each slot is 20 $\mu sec$, the propagation delay from the device to the wireless gateway is 1 $\mu sec$, and that four other wireless devices were continuously competing for the wireless media along with each wireless CPV client. With these parameters, CPV's false accepts was at 2\% and false rejects at 4\%. Although CPV's efficacy was affected by the WiFi access network, increasing the number of iterations can improve the results; see~\cite{abdou2016wireless}.

\subsection{Live Demo}

A live demo of CPV is currently running on \url{http://cpv.ccsl.carleton.ca}. This link hits a webserver in Ottawa, Canada, which enables clients to verify if they are present inside a US-based triangle determined by verifiers in San Francisco, Las Vegas, and San Diego. The verifiers are provided by hosting services DigitalOcean, ServerPoint, and M5 Hosting. Each VM has a 500MB RAM, and runs Ubuntu 16.04. NTP is used to synchronize their clocks. Additionally, each verifier issues an NTP query every 30 minutes using the \texttt{ntpq} utility to calculate the clock offset with the other two verifiers, which is added to the calculated OWDs between the verifiers for more accurate OWD estimates. Each verifier issues a timestamp to the other two verifiers every six seconds for direct OWD measurements between the verifiers.

A Java implementation of a CPV verifier runs on top of a lightweight custom-written WebSocket server, which is also implemented in Java. When a location verification request is initiated, the verifiers first check that it was issued from the authentic server (the one based in Ottawa in that demo implementation), because this server digitally signs connection IDs when they are issued. Additionally, each exchanged timestamp message between the verifiers through the client is corroborated using an MD5-based HMAC (a stronger HMAC is recommended to be used in practice). For the currently running demo, eight delay-measuring iterations are performed, once every 300ms. When all iterations are performed, the verifiers send the measured delays back to the Ottawa server, which processes the result and returns it to the browser as a jQuery response. No client puzzles are implemented yet in this demo as of this writing, nor any automatic calibration of CPV's parameters (Section~\ref{sec:calibration}). Instead, the main server has manually set parameters of $\epsilon=10ms$ and $\tau=0.7$, which are static and used across all clients.

\section{Server Location Verification (SLV)}

Analogous to CPV but on the server side, SLV~\cite{abdou2017server} works by finding evidence of a server's physical presence inside a geographic region by measuring the server's network delays. A browser typically communicates with an \textsl{SLV Manager}, which orchestrates a network of server location verifiers. The challenges faced in doing so are quite different from verifying clients: (1) clients do not normally have the ability to write and run code on the server, whereas that was easily achievable by the server on the client typically using Javascript; (2) The common physical distribution of webcontent using Content Distribution Networks (CDNs) and replication technologies begs the questions: \emph{Of the multiple physical servers that may serve client content, which such servers should be selected to geograpihcally locate (verify), in order to provide a useful server-authentication service? And how should that machine be identified?}

The answers to these questions depend on the threat model and the application for which geolocation is to be used. Since the goal of SLV is to reinforce server authentication, the implementation of SLV takes the view that the first machine that terminates the client's TCP (and TLS) handshake is the most critical one. The protection provided from verifying that first machine would be comparable to that provided by TLS, \eg in the cases where the browser fetches content from multiple machines some of which are not TLS-protected (\ie a page with mixed content).

For deciding on the mechanism used to identify machines, it is important to dissect Man-in-the-Middle (MitM) and server impersonation attacks. In MitM attacks, an adversary hijacks network traffic intended to the authentic server, and relays it to the authentic server with or without modification. Hijacking could occur on several layers of the network stack, as follows. Note that using uncompromised TLS protects against the following hijacking cases; the value of using server location to reinforce server authentication is more profound for non TLS-enabled websites or to catch attacks against the TLS system.

\begin{itemize}
\item {\bf Case 1: Attacker's machine has a different IP address than the authentic server. }In upper layers, phishing and pharming attacks are prominent traffic hijacking examples; the outbound traffic from the client has a different IP address than that of the authentic server. If the browser submits the domain name of the visited website to the SLV Manager, the Manager may resolve it to a different IP address than that seen by the browser (which could also occur benignly in the cases of CDNs). Verifying the geographic location of that IP address then becomes useless to the browser because a MitM adversary would go undetected. It is thus important to have the browser resolve a domain and submit the IP address to the SLV Manager.

\item {\bf Case 2: Attacker's machine has same IP address as authentic server. }In lower layer hijacking (such as MAC table poisoning and ARP spoofing) and in BGP spoofing, outbound traffic from the client has the same destination IP address as that of the authentic server. Such tactics are based on routing manipulation, so that traffic intended to the authentic server's IP address reaches a different ``network location" (versus geographic location), which corresponds to the attacker's machine. \end{itemize}

In comparison to upper layers, lower layer hijacking attacks tend to be more scalable, affecting larger proportion of clients. For MAC table poisoning and ARP spoofing, the closer the attacker's machine to the authentic server's network, the more the affected clients. Likewise, BGP spoofing can cause traffic hijacking at a global scale~\cite{hiran2013characterizing}. This implies that as with higher layer traffic hijacking attacks discussed above, identifying the server by its IP address will likely allow the SLV Manager to detect if the browser-intended machine is at a different geographic location than that asserted, \eg from a static location mapping previously obtained for that IP address.

Revisiting the above questions, this means that if SLV targets the IP address (as resolved by the browser) of the first machine that the client initially handshakes (regardless of whether the browser will be instructed to fetch other content from different places later in the session), it can detect most of the above server impersonation attacks. 
\begin{figure}
\centering
\includegraphics[scale=0.25]{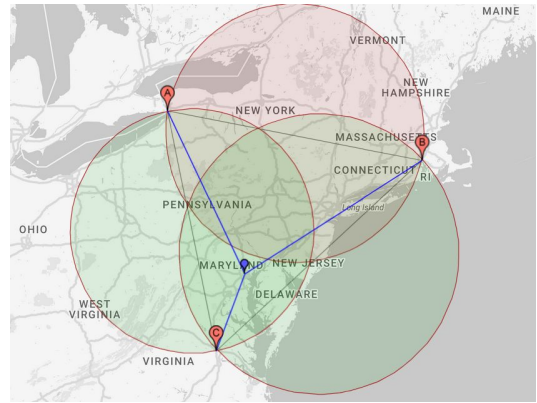}
\caption{Server Location Verification (SLV) using network measurements from three verifiers (A, B, and C) to a server. Best viewed in color. Map data: Google, INEGI.}
\label{fig:slv}
\end{figure}

\subsection{Verification Mechanism}

After obtaining an unverified server location assertion, three verifiers surrounding that location are selected. The verifiers measure network RTTs to the server over several layers, including Application using HTTP request-response times and Transport using TCP handshake responses. By means of comparing these delays with the delays between the verifiers, each pair of verifiers then verify whether the server is physically present inside the circle whose diameter is the physical distance between the verifier pair, and center is the midpoint between them (see Fig.~\ref{fig:slv}).

\subsection{Evaluation Results}

Pilot testing of $\sim$200 experiments was conducted on SLV using PlanetLab, half of which were true location assertions made by servers and the other half were false assertions. As with CPV, the rates of false rejects and false accepts were the fundamental evaluation parameters. SLV resulted in 0\% false accepts and 2.4\% false rejects~\cite{abdou2017server}. Although the false reject rate may seem high for some applications, it can be improved by proper selection of verifiers, \eg those with sufficient network bandwidth and processing resources.

\subsection{SLV Browser Extension}

We have built a Firefox browser extension to reinforce TLS by integrating the webserver's verified physical location as described above into the server authentication process. The extension sends the IP address of the server to the SLV Manager, and receives the location verification result. The extension uses \emph{FlagFox} to obtain an assertion (unverified) for the server's location. It also displays a flag in the URL (recall: Fig.~\ref{fig:flagfox}), adding to that a green tick mark or a red cross indicating whether the location asserted by FlagFox is true (according to SLV's verification) or not. This process takes a few seconds to execute, during which a throbber is displayed by the flag instead. Note that such visual cues are only meant as visual feedback (in prototypes), and not an indication that we would expect end-users to take decisions upon. See below for how policies could be implemented to automatically take decisions on behalf of users.

{\bf Server location pinning in the browser.} To avoid having the user interpret these icons, the SLV extension is supported with a \emph{location pinning} feature, whereby a browser saves that a website (identified by its URL) was previously verified to host content from this geographic location (analogous to key pinning~\cite{kranch2015upgrading}). Although location verification is preformed based on the IP address (recall: the SLV Manager only receives an IP address from the browser), location pinning in the browser is based on the domain name. Upon receiving the verification result for a website, its location gets pinned only if the result was positive, \ie the location was verified as true. This operation follows a trust on first use (TOFU) concept.

In general, for interpreting a received verification result, the SLV extension checks if that location (to some degree of geographic precision) was pinned before for that website. The result of the operation is threefold: Critical, Suspicious, or Unsuspicious. The first is when the verification result for a previously pinned location was negative. A Suspicious outcome is when location verification fails, but no location was pinned before for that website. Finally, an Unsuspicious outcome is when location verification passes for a domain that was not previously pinned. Note that these are only meant to illustrate how a client might utilize SLV, but we expect different applications would make different choices.

Such outcomes could result in the browser automatically taking decisions, \eg through a policy-based engine. An example would be to instruct the browser to block/terminate the connection for all Critical outcomes of the user's personal banking website. Such a language is subject to more research scrutiny, and is not yet part of the above SLV extension.

\section{Concluding Remarks}

This article gave a technical overview of recent advancements in the field of secure geolocation over the Internet. Two technologies were explained, CPV and SLV, to address client and server geolocation respectively. Both rely on network timing measurements for secure location verification, taking into consideration safety measures to limit adversarial manipulations. Of the wide variety of applications that may benefit from secure location information of clients and servers, reinforcing authentication (\ie location-aware authentication) for both ends remains an important example. Future research on CPV and SLV includes further enhancing their accuracy (in terms of the false reject and accept rates) and efficiency for large scale deployments in practice.

\end{document}